\documentclass[pdflatex,notitlepage,showkeys,floatfix,aps,pra]{revtex4-2}
\usepackage[utf8x]{inputenc}
\usepackage{graphicx}
\usepackage{epstopdf}
\usepackage{epsfig}
\usepackage{amssymb}
\usepackage{setspace}
\usepackage{amsmath}
\usepackage{textcomp}
\usepackage{url}
\usepackage[pdftex]{color}
\usepackage{subfigure}
\usepackage{physics}
\usepackage{dsfont}
\usepackage{dcolumn}
\usepackage[T1]{fontenc}
\usepackage{mathptmx}
\usepackage{etoolbox}
\usepackage{cancel}
\usepackage{ulem}

\begin{document}

\title[Estimating the Number of States of a Quantum System via the Rodeo Algorithm for Quantum Computation]{Estimating the Number of States of a Quantum System via the Rodeo Algorithm for Quantum Computation}

\author{J. C. S. Rocha}
  \email{jcsrocha@ufop.edu.br}
  \affiliation{Departamento de Física, ICEB, Universidade Federal de Ouro Preto, Ouro Preto, Minas Gerais, Brasil.}
\author{R. F. I. Gomes}
  \affiliation{ILACVN, Universidade Federal da Integração Latino-Americana, Foz do Iguaçu, Paraná, Brasil.}
\author{W. A. T. Nogueira}
\author{R. A. Dias}
  \affiliation{Departamento de Física, ICE, Universidade Federal de Juiz de Fora, Juiz de Fora, Minas Gerais, Brasil.}

\date{\today}

\begin{abstract}
In the realm of statistical physics, the number of states in which a system can be realized with a given energy is a key concept that bridges the microscopic and macroscopic descriptions of physical systems. For quantum systems, many approaches rely on the solution of the Schr\"odinger equation. In this work, we demonstrate how the recently developed rodeo algorithm can be utilized to determine the number of states associated with all energy levels without any prior knowledge of the eigenstates. Quantum computers, with their innate ability to address the intricacies of quantum systems, make this approach particularly promising for the study of the thermodynamics of those systems. To illustrate the procedure's effectiveness, we apply it to compute the number of states of the 1D transverse-field Ising model and, consequently, its specific heat, proving the reliability of the method presented here.
\end{abstract}


\maketitle

\section{\label{sec:intro}Introduction} 

In 1902, J. Willard Gibbs introduced the concept of a statistical ensemble as a theoretical construct representing the entire set of possible states that a real physical system could occupy under specific conditions~\cite{gibbs}. It is widely recognized that selecting the appropriate ensemble is essential for the specific physical context, given that the equivalence between different ensembles may not be valid in certain situations~\cite{microcanonical}. This becomes particularly evident in systems characterized by long-range interactions~\cite{barre}, as exemplified in the context of gravity~\cite{votyakov}, as well as in finite systems comprising a limited number of particles~\cite{bertoldi,miranda}. In recent times, numerous problems confronted in the domains of science and technology are observed to manifest as phenomena intricately associated with diminutive dimensions~\cite{puglisi}.

The description of isolated quantum systems is appropriately achieved through the use of the quantum microcanonical ensemble~\cite{griffiths,dorje,helrich_book}. The importance of this ensemble resides in allowing us to understand how quantum mechanical effects influence the nature of thermodynamic phenomena~\cite{brody}. Furthermore, it plays a crucial role in the study of the thermodynamics~\cite{abraham} and the behavior of quantum systems at the microscopic level
~\cite{bobak}. In this ensemble, once the microstates are defined, the thermodynamic behavior of the system is promptly available via the entropy,
\begin{equation}
    S(E) = k_B\ln{\Omega(E)},
\end{equation}
where $\Omega(E)$ stands for the number of states (NoS) with energy $E$ and $k_B$ represents the Boltzmann constant~\cite{Planck}.

Conversely, there may be an interest in investigating the thermodynamic characteristics of systems that interact with their surroundings, such as those in thermal contact with a heat bath. In such scenarios, an appropriate analysis is conducted using the canonical ensemble~\cite{gibbs}. In this method, the fundamental quantity is the partition function, which can be regarded as the Laplace transform of the NoS~\cite{luscombe2021statistical}. This relation can be mathematically expressed as
\begin{equation}
Z(\mathfrak{B}) = \int \Omega(E)e^{-\mathfrak{B} E} \mathrm{d}E,
\label{eq:partition}
\end{equation}
where $\mathfrak{B} = \beta + i b$ represents the complex inverse temperature, with $\beta = 1/k_BT$ denoting the standard canonical inverse temperature. The canonical ensemble establishes a connection to thermodynamics through the Helmholtz free energy,
\begin{equation}
F(\mathfrak{B}) = -\frac{1}{\mathfrak{B}} \ln{Z(\mathfrak{B})}.
\label{eq:free_energy}
\end{equation}
Although the complex temperature has no physical meaning, the analytic continuation of the thermodynamic potentials can unveil certain properties of physical systems as $b \to 0$~\cite{Lee_Yang, Fisher, Rocha_2024}.

Hence, the determination of the NoS emerges as the quintessential procedure in statistical mechanics, aiming to encapsulate the whole thermodynamic information related to the physical systems and elucidate its thermodynamic behavior across various physical contexts.

More specifically, in the realm of quantum systems, solving the time-independent Schr\"odinger equation constitutes a fundamental component of many approaches to address this estimation. When analytical solutions are impractical, employing measurements on a virtual model of the system can be a suitable alternative. In this scenario, quantum computation presents a natural approach to handle this challenge, since it can account for purely quantum phenomena including entanglement states~\cite{feynman,lloyd,abrams,buluta,brown,georgescu}.

Recently, Choi et al.~\cite{RodeoPRL127} introduced the rodeo algorithm as a promising technique for preparing states of quantum many-body system models on quantum computers. This algorithm is capable of preparing the eigenstates of any observable. Particularly Hamiltonian eigenstates, thereby addressing solutions to the time-independent Schr\"odinger equation. For this estimation, the algorithm requires a hypothesized solution to the eigenproblem as input. According to the authors, a crucial requirement of this approach is a significant overlap between the wave function of the input state and that of the desired eigenstate. They propose several strategies for determining the initial state: setting it based on physical intuition regarding the nature of the wave functions, using the tensor product of eigenstates from smaller subsystems, or employing the quantum approximate optimization algorithm~\cite{RodeoPRL127, farhi2014quantum}. By scanning the energy domain, i.e. repeatedly running the algorithm with target energies within the range $[E_{min}, E_{max}]$, this method can also output the energy spectrum of the system. 

In this work, we propose a protocol for analyzing the algorithm's output, leading to an alternative strategy for inputting the target state specifically to estimate the NoS. This protocol enhances the capability to determine the probability of measuring the input state with a given energy $E$. Moreover, we show how to use this feature to compute the NoS by scanning the energy domain for all states of the computational basis.  This general methodology circumvents the need for intricate state space exploration to prepare initial states. Unfortunately, the number of states in the basis increases exponentially with the system size. Specifically, for a two-level system, the basis consists of $2^M$ states, where $M$ is the number of particles. Since the energy scanning of different states is independent, they can be performed in parallel. Thus, the method's performance is constrained by the availability of qubits rather than by time requirements, thereby this limitation can be mitigated with the development of large-scale quantum computers. 
We firmly believe that a suitable alternative to address this issue is the development of a Monte Carlo algorithm, based on the principles outlined here, to approach the NoS by sampling a representative small subset of the basis states.

We demonstrate the methodology's effectiveness by investigating the 1D transverse-field Ising model. Our choice of this model is motivated not only by its suitability for a direct comparison with exact results but also, despite certain constraints~\cite{gu}, by its ability to reformulate a given problem in terms of pairwise interactions among Ising spins~\cite{lucas,ortiz}. For instance, the quadratic unconstrained binary optimization encoding, recognized for its close association with the Ising energy function, currently emerges as a fruitful modeling strategy in quantum computation approaches. This includes applications in quantum artificial intelligence~\cite{neven} and soft matter physics~\cite{slongo,micheletti}. Therefore, this establishes a framework for future investigations.

This paper is structured as follows: In subsection~\ref{subsec.rodeo}, we provide a summary of the rodeo algorithm, after that, in subsection~\ref{subsec.nos}, we demonstrate how it can be used for estimating the NoS. In subsection~\ref{subsec:error} we discuss the sources of errors in these estimations and strategies for managing them. Section~\ref{sec:results} showcases results for the 1D transverse-field Ising model. Finally, Section~\ref{sec:conclusions} outlines our conclusions and presents perspectives for future work.

\section{\label{sec.method}Methodology} 
\subsection{\label{subsec.rodeo}The Rodeo Algorithm} 

\begin{figure}[!ht]
\includegraphics[scale=0.48]{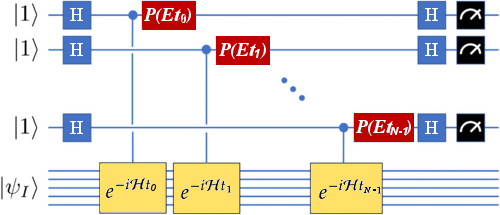}
\caption{Circuit diagram for the rodeo algorithm. The system starts in an arbitrary state $|\psi_I⟩$. The ancillary qubits are initialized in the state $|1⟩$ and operated on by a Hadamard gate $\mathrm{H}$. We use each ancilla denoted by $k=0,\cdots, N-1$ to control the time evolution of the 
Hamiltonian ($\mathcal{H}$) for a random time $t_k$. This is followed by a phase rotation $P(E t_k)$  on each one, another Hadamard gate $\mathrm{H}$, and a measurement on the computational basis.}
\label{fig:RodeoPRL}
\end{figure}

Similar to other quantum computing algorithms~\cite{Kitaev, Shor, Simon, Deutsch-Jozsa}, the rodeo algorithm exploits the phenomenon of phase kickback. This phenomenon refers to the fact that controlled operations affect both the target and control qubits, with these effects corresponding to phase shift operations. Specifically, during a controlled operation, the phase shift resulting from applying a unitary operator to one of its own eigenstates is effectively transferred to the control qubit, which is initially set in a superposition of the basis states $\ket{0}$ and $\ket{1}$~\cite{Cleve, ciaran, nagata}.
With this in mind, the key steps of the rodeo algorithm can be outlined as follows:

\begin{enumerate}

\item {\bfseries Initialization:} 

\begin{enumerate}

\item  {\bfseries  Target state $\ket{\psi_I}$:} \\ {\footnotesize *This state serves as an initial guess for one of the Hamiltonian eigenstates.}

\begin{enumerate}
\item Allocate a set of $M$ qubits required to represent the target system. 
\item Prepare these qubits into the desired initial state.
{\small  \begin{itemize} \item As discussed in the introduction, there are several strategies for the preparation of the target state. In this work, we prepare them in one of the basis states, $\ket{n}$, where $n$ represents a non-negative integer. The state notation with integers denotes the binary representation of that integer; for instance, $\ket{5} = \ket{0\cdots 0101}$. Typically, the qubits are initially set in the state $\ket{0}$. If that is the case, the state preparation is performed by applying the NOT operator, $X$, on the qubits that need to be set to $\ket{1}$, then
\[
\ket{\psi_1} = \ket{n}  = \bigotimes_{i=0}^{M-1} \ket{s_i} = \ket{s_0 s_1 \cdots s_{M-1}}, \quad \text{where } \ s_i = 0 \ \text{or} \ 1.
\]
 \end{itemize}}
\end{enumerate}

\item {\bfseries Ancillary Qubits:} 
\begin{enumerate}
\item Allocate another set of $N$ qubits to serve as auxiliaries.
\item Prepare each of then on the state $\ket{-} = (\ket{0}-\ket{1})/\sqrt{2}$.
{\small  \begin{itemize} \item  If the qubit is initially set in state $\ket{0}$, this step can be accomplished by applying the NOT operator, $X\ket{0} = \ket{1}$, followed by the Hadamard gate, $H\ket{1} = \ket{-}$, in each ancilla qubit. \end{itemize}}
\end{enumerate}
\end{enumerate}

\item {\bfseries Rodeo Operation:} 
\begin{enumerate}
 \item {\bfseries Controlled Time Evolution Operation:} Apply the controlled version of the time evolution operator $\mathcal{O} = \exp(-i\mathcal{H} t_k)$ to the target state. The $k$-th ancilla qubit controls this operator.
 {\small \begin{itemize}
     \item Here $\mathcal{H}$ stands for the Hamiltonian and $t_k$  denotes a random time interval drawn from a Gaussian distribution.
 \item If the state of the target qubits $\ket{\psi_I}$ is an eigenstate of $\mathcal{H}$, this step effectively transfers the system's evolution to the $k$-th ancilla qubit. 
 \end{itemize}}
 
 \item {\bfseries Phase Shift Operation:}  Apply the phase shift operator $P(\phi)$ with $\phi = Et_k$ to the $k$-th  ancilla qubit.
 {\small \begin{itemize}
 \item $E$ is a speculated guess for the eigenvalue associated with $\ket{\psi_I}$. 
 \item If $\mathcal{H}\ket{\psi_I} = E\ket{\psi_I}$, the phase shift operator will precisely counteract the effect of the time evolution operator. This occurs because these operators shift the ancilla qubit by the same phase but in opposite directions, resulting in this qubit returning to its initial state.
 \end{itemize}}
 
 \item {\bfseries Loop:} Repeat for $k$ ranging over the $N$ ancillary qubits.
\end{enumerate}

\item {\bfseries Measurement:} 
\begin{enumerate}

\item {\bfseries Change of basis:} Apply another sequence of Hadamard gates to the ancillary qubits.
\item {\bfseries Measurement:} Perform measurements on the computational basis for this set.
{\small \begin{itemize} \item If all ancillary qubits are observed to be in the state $\ket{1}$, i.e. in their initial state, it indicates that the time-independent Schr\"odinger equation was successfully solved. \end{itemize}}
\end{enumerate}

\end{enumerate}
The core idea here is exploiting the phase kickback phenomenon such that, if the inputs $\ket{\psi_I}$ and $E$ satisfy the time-independent Schr\"odinger equation $\mathcal{H}\ket{\psi_I} = E\ket{\psi_I}$, the phase shift will counteract the controlled time-evolution operator, resulting in the ancillary qubits being found in their initial state at the end of the process. The circuit diagram for the rodeo algorithm is shown in Fig.~\ref{fig:RodeoPRL}. 

\subsection{\label{subsec.nos}Estimating the Number of States} 
A substantial number of cycles with distinct time intervals $t_k$ is demanded to ensure the solution's reliability of the rodeo Algorithm. This can be achieved by either employing numerous ancillary qubits - which is currently impractical - or by repeating the process a suitable number of times. We denoted the number of the algorithm repetitions as the number of rounds, $N_{\text{rounds}}$.

In this sense, we want to introduce an alternative strategy to analyze the rodeo algorithm.  Rather than linking a specific state of the ancillary qubits to the solution of the Schr\"odinger equation, we suggest binding the average of the measurements of those qubits with the NoS. We refer to this average as the Score Average (SA) of the rodeo algorithm, for given inputs $\ket{\psi_I}$ and $E$, it is computed by the formula: 
\begin{equation}
    \overline{h}(E,{\psi_I}) = \frac{1}{N N_{\text{rounds}}}\sum_{i=1}^{N_{\text{rounds}}}\sum_{k=0}^{N-1} \langle \sigma^z \rangle_{k,i},
    \label{eq:hrodeo}
\end{equation}
where $\langle \sigma^z \rangle_{k,i}$ is the expectation value of the measurement of the $k^{th}$ ancilla qubit in the $i^{th}$ round. We reserve the overbar to designate the average over a sample set, while angle brackets are used for the quantum expectation value. It is worth noting that the angle brackets notation implicitly accounts for multiple algorithm cycles (with fixed $t_k$), referred to as the number of shots, $N_{shots}$, to calculate the quantum expectation values. In this work, $N_{shots}=1000$.

One should note that the ancillary qubits are twisted side-to-side by angles proportional to $t_k$. If the time interval $t_k$ is drawn from the Gaussian distribution (as originally proposed) the SA should behave like
 \begin{equation} \label{eq:get}
     -\overline{h}(E,{\psi_I}) =  \sum_{x} |\bra{x}\ket{\psi_I}|^{2}   e^{-\frac{d^2\left[ (E - E_{x}) \right]^2}{2}}\cos{\left[ (E - E_{x})\tau \right]},
\end{equation}
where $\tau$ stands for the mean value and $d$ stands for the standard deviation of the Gaussian distribution. This relation is deduced in the Appendix~\ref{appendix_SA}. From this point onward, we will represent the eigenstates of the Hamiltonian as $\ket{x}$ and their associated eigenvalues as $E_x$. Thus, the summation index $x$ signifies that it ranges over all the $2^M$ eigenstates, including degenerescence. 

It is well known that $\ket{x}$ can be expressed as a linear combination of the basis states, i.e.
\begin{equation}
    \ket{x} = \sum_{n'=0}^{2^M-1} c_{x,n'} \ket{n'},
    \label{eq.eigen_n}
\end{equation}
where $c_{x,n'}$ denotes the $n'$-th coefficient associated with the eigenstate $\ket{x}$. By selecting one state from the computational basis as the input, $\ket{\psi_I}= \ket{n}$, and incorporating it along with Eq.~(\ref{eq.eigen_n}) into  Eq.~(\ref{eq:get}), due to the orthonormality of the computational basis, it results in
 \begin{equation}
    -\overline{h}(E,n) = \sum_{x}
 c_{x,n}^2 e^{-\frac{d^2\left[ (E - E_{x}) \right]^2}{2}}\cos{\left[ (E - E_{x})\tau \right]}.
 \label{eq:SAbehavior}
 \end{equation} 
In accordance with the normalization condition for the eigenstates, i.e.
\begin{equation}
 |\bra{x}\ket{x}|^2 = 1 \  \ \Rightarrow \ \ \sum_{n=0}^{2^M-1} c_{x,n}^2 =1,
\end{equation}
it can be stated that summing Eq.~(\ref{eq:SAbehavior}) over all basis states leads to:
 \begin{equation} 
    \sum_{n=0}^{2^M-1} -\overline{h}(E,n) = \sum_{x}
 e^{-\frac{d^2\left[ (E - E_{x}) \right]^2}{2}}\cos{\left[ (E - E_{x})\tau \right]}.
 \label{eq:sumSA}
 \end{equation} 
By choosing $d\gg 1$ and $\tau \ll 1$, Eq.~(\ref{eq:sumSA}) can be approximated as:
 \begin{equation} \label{eq.approach_H}
    \sum_{n=0}^{2^M-1} -\overline{h}(E,{n}) \approx \sum_{x} \delta[E-E_x],   
 \end{equation} 
where
\begin{equation}
    \delta[E-E_{x}] = \left\{\begin{array}{cc}
         1 & \text{if } E = E_x \\
         0 &  \text{if } E \ne E_x 
    \end{array}\right. ,
\end{equation}
is the Kronecker delta. Note that the right term of Eq.~(\ref{eq.approach_H}) can be seen as the number of degenerate eigenstates with energy $E$.
Thus, the NoS can be ascertained by implementing the rodeo algorithm with each basis state as input, subsequently aggregating all SA's obtained, i.e:
 \begin{equation}
\Omega(E)\approx \sum_{n=0}^{2^M-1} -\overline{h}(E,{n}).
    \label{eq:nos_simulated}
\end{equation}
The power of this definition lies in the fact that we can assess the NoS without any knowledge of the eigenstates. Besides that, the SA can be calculated separately for each input, offering a straightforward parallel framework to explore the simulation.

\subsection{\label{subsec:error}Error analysis}
We have adopted a perspective in which no prior knowledge of the system is available. Hence, we compute the SA for each basis state, considering discrete energy values within a specified range of interest, $[E_0, E_f]$, with step sizes set to $\varepsilon$.  More specifically, the energy of the $\ell$-th step is equal to $E_{\ell} = E_0 + \ell \varepsilon$. As the eigenvalue can be situated between two of these energy bins, the precision of its estimation is proportional to $\varepsilon$.

In turn, the Trotterization process - employed to address the time evolution of Hamiltonians containing non-commutative terms - is also an inherent source of error. The $m$-order decomposition of the time evolution operator of a Hamiltonian consisting of $\Gamma$ terms, i.e. $\mathcal{H} =\sum_{j=1}^{\Gamma} \mathcal{H}_j$, can be written by the recursive relation:
\[\begin{array}{ll}
     \mathcal{S}_1(t_k) = &e^{-i\mathcal{H}_1 t_k} \cdots e^{-i\mathcal{H}_{\Gamma}t_k}, \\
     \mathcal{S}_2(t_k)  =  &e^{-i \mathcal{H}_1\frac{t_k}{2}}\cdots e^{-i\mathcal{H}_{\Gamma}\frac{t_k}{2}} e^{-i\mathcal{H}_{\Gamma}\frac{t_k}{2}} \cdots e^{-i\mathcal{H}_1\frac{t_k}{2}}, \\
     \ \vdots & \\
     \mathcal{S}_{m}(t_k)  =& \mathcal{S}^2_{m-2}(p_m t_k)\mathcal{S}_{m-2}((1 - 4p_m) t_k)\mathcal{S}^2_{m-2}(p_m t_k),
\end{array}
\]
where $p_m = 1/(4 - \sqrt[1-m]{4})$~\cite{childs}.
The Suzuki-Trotter approach is then defined as:
\begin{equation}
    e^{-i\mathcal{H}t_k} = \left(  \mathcal{S}_m\left(\frac{t_k}{r}\right) \right)^r + \mathcal{O}\left( \frac{ t_k^{m+1}}{r^m} \right),
\end{equation}
where $r$ is the number of Trotter steps~\cite{suzuki-91}. Hence, to stipulate a precision $\delta$ in the time-evolution operation we must set the Trotter-number as
\begin{equation}
    r = \mathcal{O}\left( \frac{t_k^{1 + \frac{1}{m}}}{\delta^{\frac{1}{m}}}\right).
    \label{eq:ntrotter}
\end{equation}
However, this pursuit is subject to a compromise between error minimization and computational time. Furthermore, opting for large values of $m$ and $r$ may lead to cumulative errors arising from the repeated application of quantum gates. 

Given that $t_k$ is randomly selected, it is imperative to consider the influence of the parameters associated with the probability density function on error estimation. Specifically, since $d$ governs the decay of the peaks in the SA, as seen in Eq.~(\ref{eq:get}), it directly impacts the accuracy of the NoS for eigenvalues situated between two energy bins. Because the values considered in the energy discretization might reside in the tail of this Gaussian decay, potentially leading to an underestimation of $\Omega(E)$. Moreover, increasing the values of $d$ implies drawing larger time intervals more frequently, resulting in a higher frequency of larger values for the Trotter numbers. In turn, large $\tau$ leads to large time intervals on average. These extended time intervals not only imply increased computational effort but also lead to associated errors, both induced by the Trotterization process, as mentioned in the previous paragraph

Since the SA is given by an average value, Eq.~(\ref{eq:hrodeo}), a statistical fluctuation must also be considered. This is evaluated by the sample standard deviation,
\begin{equation}
s({h}(E,{\psi_I})|t) = \sqrt{\frac{\mathrm{Var}\big({h}(E,{\psi_I}|t)\big)}{N N_{\text{rounds}}}},
\label{eq:sde}
\end{equation}
where $\mathrm{Var}(h) = \overline{\big(h^2\big)} - \Big(\overline{h}\Big)^2$ is the variance~\cite{kenney}.  Thus, the error can be mitigated by considering large values for $ N_{\text{rounds}}$. Once again, we confronted the trade-off between minimizing errors and computational time.

Furthermore, since each ancilla is attached to three independent transformations (two Hadamard gates and one Phase gate) and the controlled operation on the target system, their addition expands the circuit's width and depth, which are respectively defined as the number of qubits required to execute the algorithm and the number of gates that can be carried out simultaneously \cite{barenco1995}. Therefore, implementing the algorithm with additional ancillary qubits increases the spatial and temporal complexity of the circuit, which depends respectively on the total number of qubits and the time required to perform these operations~\cite{bernstein1997, bouland2018}.
Moreover, considering that the decoherence process is inherent to any qubit maintained in a realistic environment \cite{nielsen2010quantum}, the entire system becomes more susceptible to noise when the algorithm is applied to real devices. Consequently, this susceptibility results in a loss of information, thereby decreasing the total success rate for $-\overline{h}(E,{\psi_I})$ calculated through the simulations.
Finally, like any other quantum algorithm relying on the coherence of quantum states, it is essential to consider quantum error correction methods to protect the purity of qubits \cite{jaeger2006quantum} on real quantum computers, reducing the interaction with external environments and other sources of quantum noises~\cite{devitt}.
\section{\label{sec:results}Results} 
In this paper, we illustrated the estimation of $\Omega(E)$ via the rodeo algorithm by studying the 1D transverse-field Ising model. The Hamiltonian of this model was introduced by P.G.~de~Gennes in 1963 
as a description for hydrogen-bonded ferroelectric~\cite{degenes}, but its structure also provides a useful representation for many other systems, in particular, insulating magnets. Moreover, R.B.~Stinchcombe has shown a list of physical systems that can be described by this model~\cite{stinchcombe}, which can mathematically be expressed as
\begin{equation}
    \mathcal{H} = -J\sum_{<i,j>} \sigma^z_i\sigma^z_j - B\sum_i \sigma^x_i.
\label{eq:ising}
\end{equation}
For the present work, the parameters in Eq.~(\ref{eq:ising}) were defined as
$J=1$ stands for the exchange constant, the symbol $<i,j>$ means that the summation was taken over the first nearest neighbors,  $\sigma^z$ ($\sigma^x$) is the Pauli-$z$ (Pauli-$x$) operators, and $B\ge0$ stands for the magnetic field strength. Besides that, we consider periodic boundary conditions. In our chosen unit system, we set fundamental constants to unity, specifically, the ones that represent the Planck constant ($\hbar = 1$), the Boltzmann constant ($k_B=1$), the Bohr magneton ($\mu_B = 1$), and the square of the spin modulus ($\sigma^2 = 1$).

\begin{figure}[!ht]
\begin{tabular}{c c}
 \includegraphics[scale=0.3]{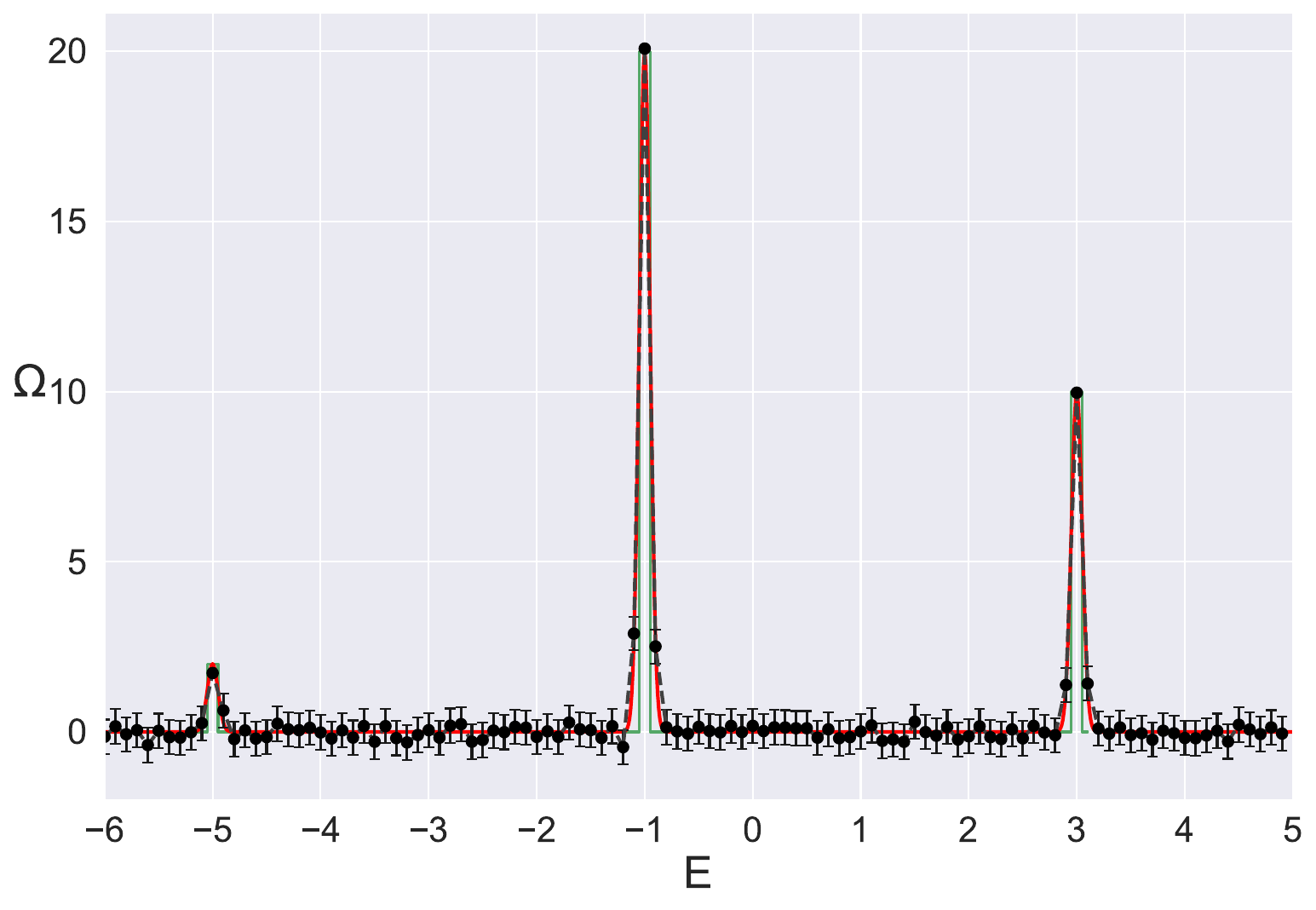} & 
 \includegraphics[scale=0.3]{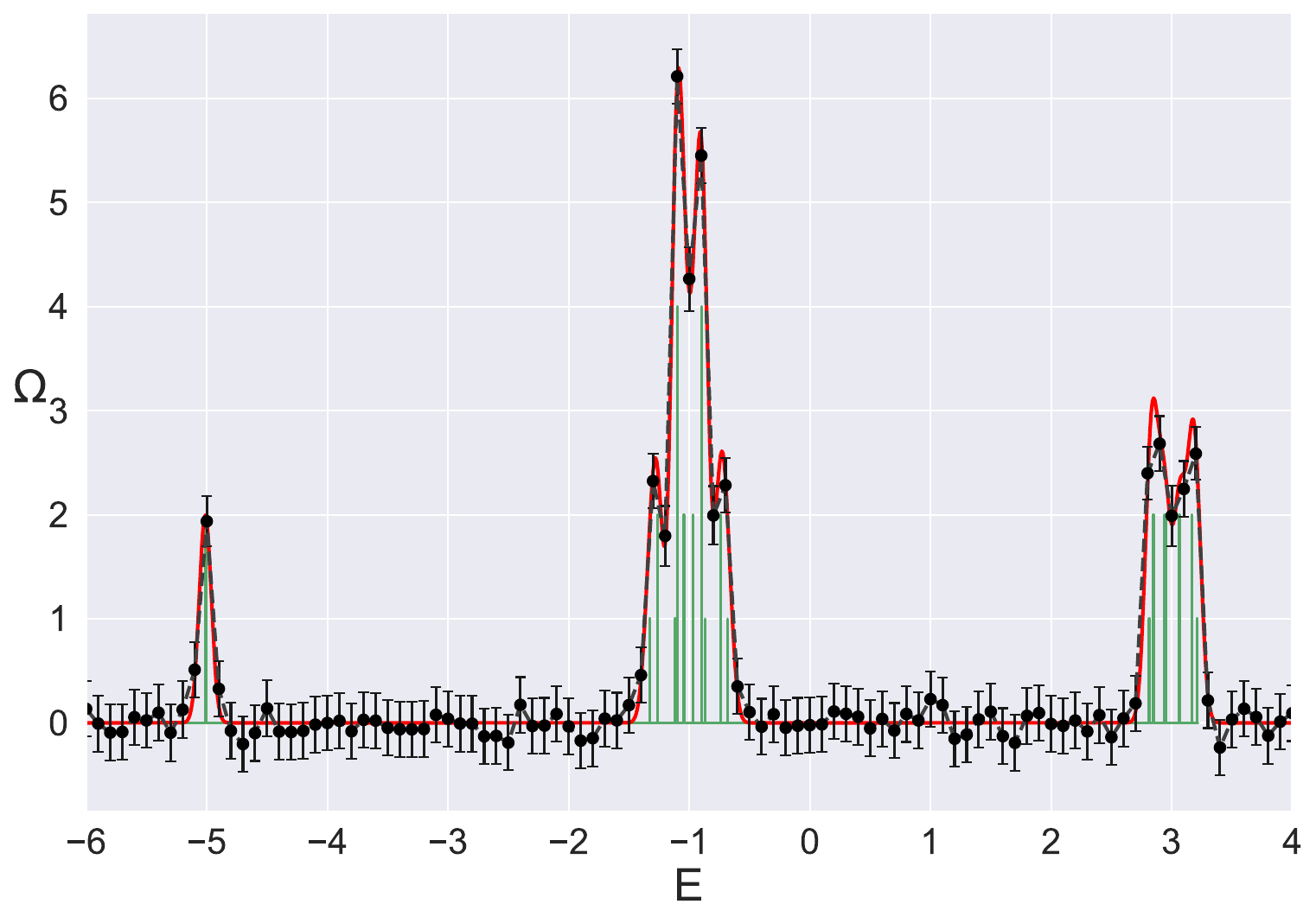} \\
 (a) & (b) \\
\includegraphics[scale=0.3]{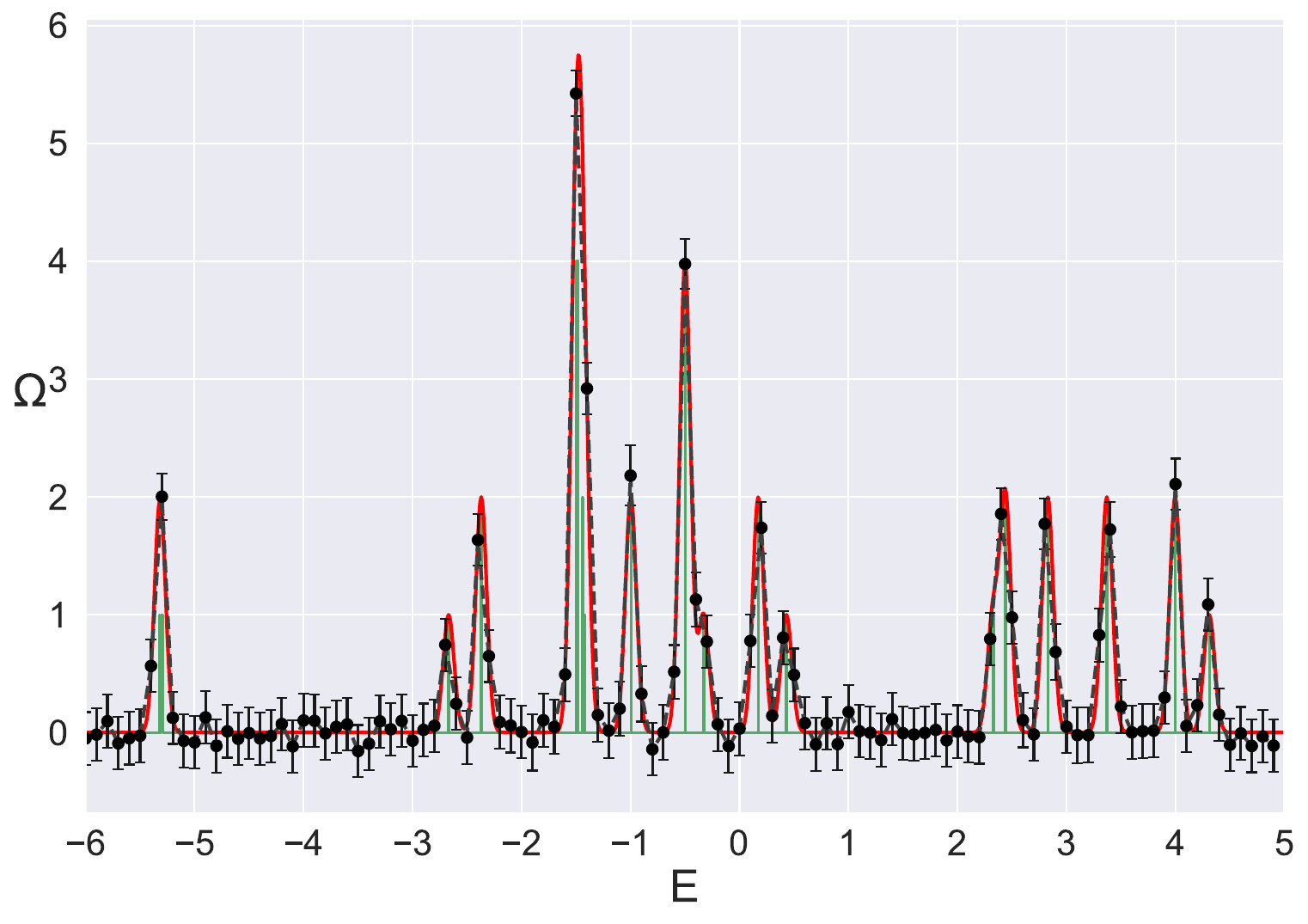} &
 \includegraphics[scale=0.3]{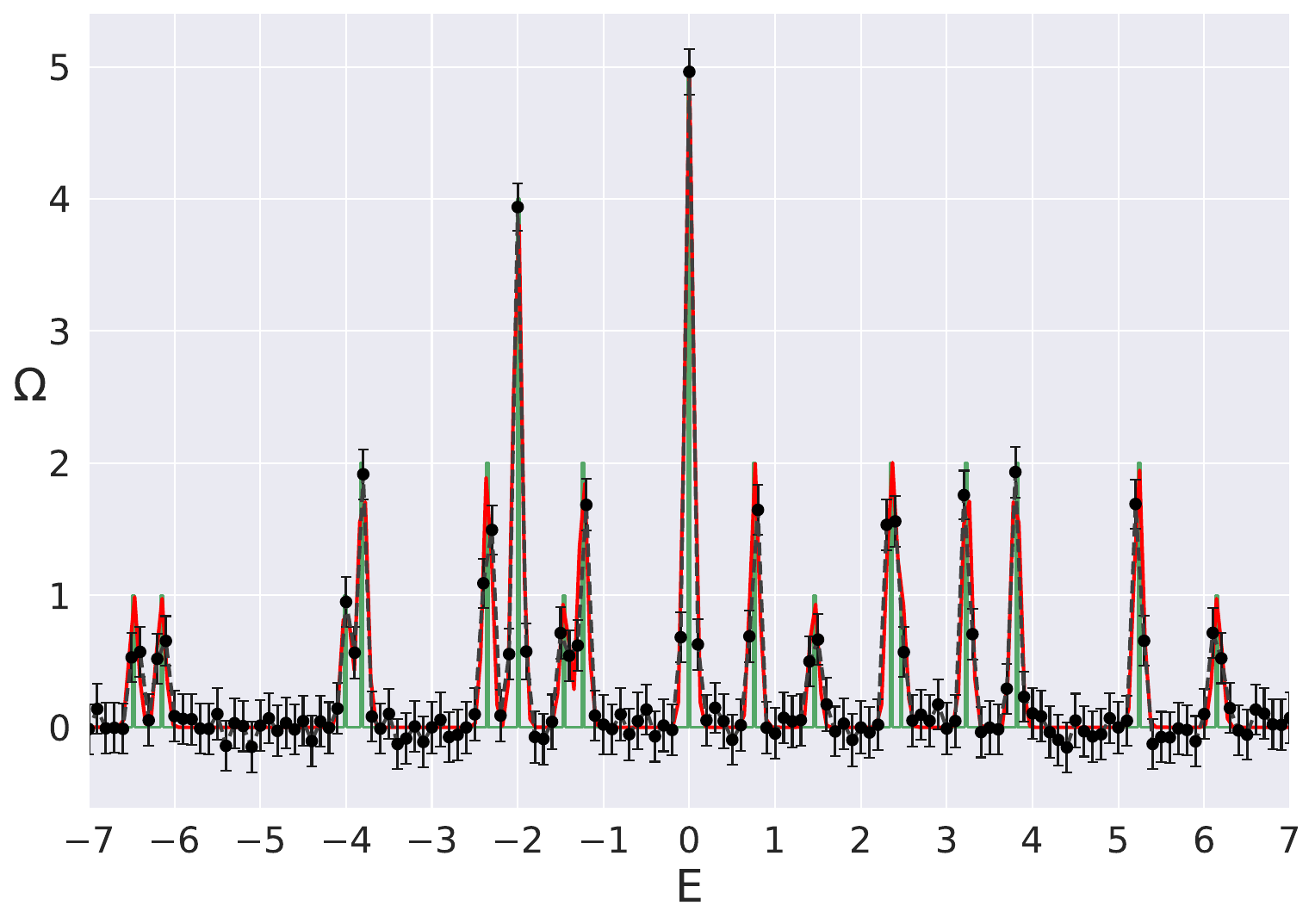} \\
 (c) & (d)
\end{tabular}
\caption{Number of states for the 1D transverse-field Ising Model with $5$ spins. Panel (a) shows the results for $B=0$, panel (b) for $B/J=0.1$, panel (c) for $B/J=0.5$, and panel (d) for $B/J=1.0$. The results obtained from the rodeo algorithm are shown in black dots, the red curve exhibits the theoretical behavior, and the green histogram represents the exact result. The algorithm parameters were selected as follows: $N=1$, $\varepsilon=0.1$, $m=1$, $\delta=0.1$, $\tau=0$, $d=20$ and $N_{\text{rounds}} = 500$.}
\label{fig:result}
\end{figure}

All simulations were conducted on systems comprising $M=5$ spins through the Pennylane quantum simulator~\cite{pennylane}, whose 
 results are shown in Fig.~\ref{fig:result} for magnetic field strengths $B=0$, $B=J/10$, $B=J/2$, and $B=J$, respectively from panels (a) to (d). Here we consider $N=1$ ancilla qubit, the energy range $[-6,5]$, $\varepsilon=0.1$, $m=1$, $\delta=0.1$, $\tau=0$, $d=20$ and $N_{\text{rounds}}=500$. 
The black dots exhibit outcomes acquired by the rodeo algorithm, i.e. evaluation of Eq.~(\ref{eq:nos_simulated}) with the SA given by Eq.~(\ref{eq:hrodeo}), and the error bars represent the standard deviation relative to this average.  The red curve shows the theoretical behavior given by Eq.~(\ref{eq:sumSA}), where $E_x$ were obtained by diagonalizing the Hamiltonian. Concurrently, the NoS reckoned up through this procedure is delineated in each panel by the green histogram in this figure.

For $B=0$, the NoS enumeration becomes straightforward through a direct evaluation of the Hamiltonian for all $32$ basis states. Since each of these states is an eigenstate of $\sigma^z$, they are inherently eigenstates of the Hamiltonian in this specific scenario. For the ground state $E=-5$, there are two distinct solutions resulting in $\Omega(-5) = 2$. The first excited states at $E=-1$ exhibits $\Omega(-1) = 20$, while the final energy level at $E=3$ holds $\Omega(3)=10$ states. Fig.~\ref{fig:result}~(a) demonstrates the successful estimation of these values by the rodeo algorithm. 

The magnetic field's influence in breaking degeneracy and its role in dictating the energy gap between split states are well-established~\cite{zeeman,griffiths_introduction_2018}. The emergence of state splitting becomes progressively evident for non-zero values of $B$. At a relatively small magnetic field strength, $B=J/10$, a perceptible dispersion of states is observed around the previously highly degenerate levels that were counted for $B=0$, see Fig.~\ref{fig:result}~(b). In this scenario, given the specified parameters, the energy gap of the split states is smaller than the decay of the exponential term in Eq.~(\ref{eq:sumSA}). As a result, there is a superposition of these terms, resulting in a smooth estimation of the NoS. 

\begin{figure}[!ht]
\includegraphics[scale=0.3]{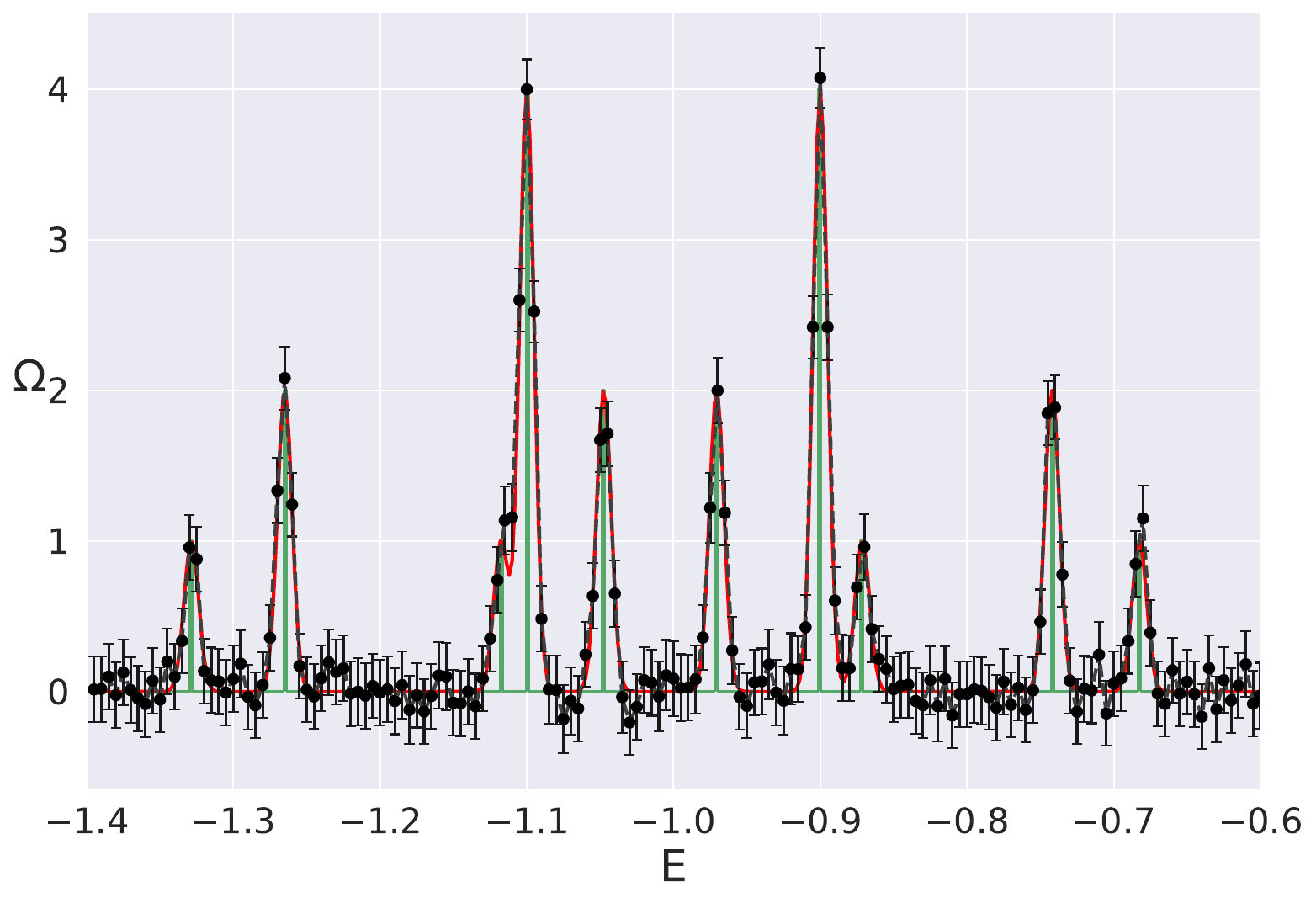}
\caption{Refinement of the central peak of the NoS for the $5$-spins transverse-field Ising model with filed strength $B/J=0.1$. The results obtained from the rodeo algorithm are shown in black dots, the red curve exhibits the theoretical behavior, and the green histogram represents the exact result. The algorithm parameters were selected as follows: $N=1$, $\varepsilon=0.005$, $m=1$, $\delta=0.1$, $\tau=0$, $d=200$ and $N_{\text{rounds}}=500$.}
\label{fig:refinement}
\end{figure}

To enhance the precision of the rodeo algorithm, the standard deviation parameter of the Gaussian distribution has to be increased. Notably, this behavior represents an intrinsic characteristic of the algorithm, which requires a refinement of the study by exploring diverse parameters of the Gaussian distribution in specific regions of maximum values.  Moreover, as mentioned before, a faster exponential decay linked to large values of $\varepsilon$ may correspond to potentially lower accuracy. As a result, the refinement process should take into account a reduction in the energy step to address this dependency. In Fig.~\ref{fig:refinement} we show the refinement of the results for $B=J/10$ in the energy range $[-1.4,-0.6]$. Here we consider $\varepsilon = 0.005$, $d=200$, and we kept $N=1$, $m=1$, $\delta=0.1$, $\tau=0$ and $N_{\text{rounds}}=500$. As expected, the NoS estimation in this region is enhanced.

As the magnetic field strength reaches a moderate level, $B=J/2$, a discernible spread of the split states across a wide energy range becomes observable, see Fig.~\ref{fig:result}~(c).
The dispersion of energy levels and, wherefore, the flatness of the NoS, result in a high transition probability to excited states. Consequently, it is highly improbable to observe large systems in an ordered state. This argument is a pictorial and qualitative standpoint of the observed quantum phase transition ($T=0$) that occurs at the critical magnetic field strength $B_C=J/2$, marking an ordered phase for $B<B_C$ and a disordered one for $B>B_C$~\cite{pfeuty}. 

However, it is fundamental to acknowledge that the non-analytic points on the thermodynamics functions that are associated with the phase transitions occur only in the thermodynamic limit~\cite{critical}.
Given the impracticality of simulating infinitely large systems, the procedure to determine critical quantities requires a finite-size scaling analysis~\cite{fss}, which is not addressed here since the primary focus of this manuscript is to introduce and elucidate the technique rather than explore the properties of any particular system.  

 Fig.~\ref{fig:result}~(d) presents the result for $B=J$ representing the case for a magnetic field significantly larger than $B_C$. This result shows an accentuated spread of the split states, making the NoS even flatter and reducing the energy gap between energy levels.

The main reason to estimate the NoS is to obtain the thermodynamic behavior of physical systems. All thermodynamic equations can be written as derivatives of a fundamental thermodynamic relation. For instance, the specific heat can be written as the second derivative of the Helmholtz free energy, Eq.~({\ref{eq:free_energy}}), with respect to the inverse temperature, i.e.
\begin{equation}
    c_B = \frac{\beta}{M} \left( \frac{\partial^2 F}{\partial \beta^2} \right)_B,
\end{equation}
whose result is promptly available once the NoS is known. 

\begin{figure}[!ht]
\includegraphics[scale=0.3]{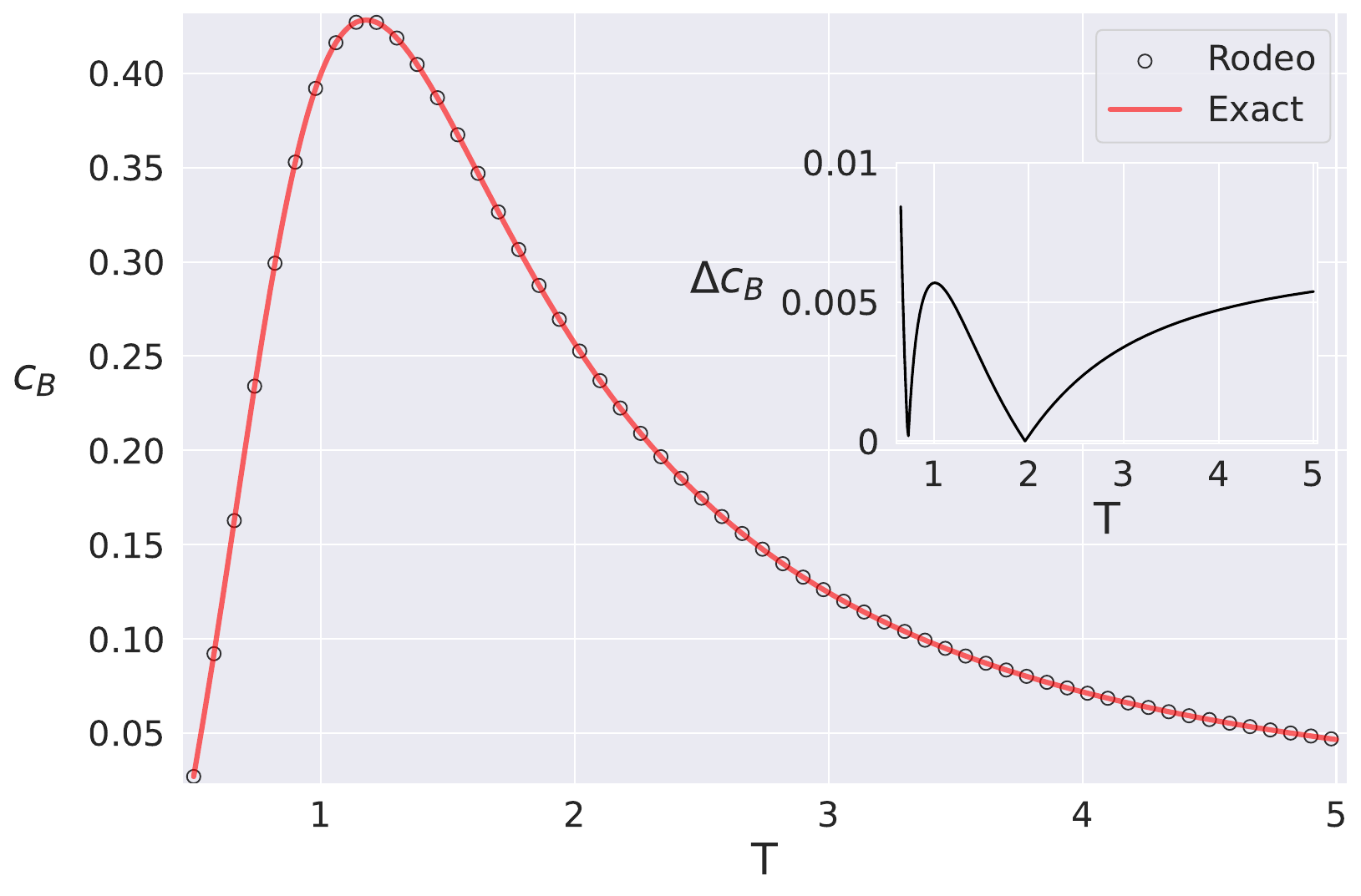}
\caption{Specific heat at a constant magnetic field $B=J/2$ of the 5 spins transverse-field Ising model via rodeo algorithm. The inset shows the relative difference between the rodeo estimate and the exact value.}
\label{fig:specificheat}
\end{figure}

The specific heat for this model at a constant magnetic field $B=J/2$ is presented in Fig.~\ref{fig:specificheat}. The black circles indicate the values obtained from the NoS estimated by the rodeo algorithm, where all points that are shown in Fig.~\ref{fig:result}~(c) were considered. The red curve represents $c_B$ obtained from the exact value of the NoS. In the inset of this graph, we show the relative difference between those evaluations, i.e.
\begin{equation}
    \Delta c_B = \left| 1 - \frac{c_{B_{\text{rodeo}}}}{c_{B_{\text{exact}}}} \right|.
\end{equation}
The rodeo algorithm estimative differs by less than $1\%$ of the exact values in the considered temperature range. Hence, it provides a piece of convincing evidence for the reliability of the method presented here.

\section{\label{sec:conclusions}Conclusions and Perspectives} 
In summary, the rodeo algorithm provides a means to estimate the number of states for all energy levels of a quantum system by averaging the measurements of the ancillary qubits, whose value is named the Score Average. In addition, no prior knowledge of the energy eigenstates is needed. We successfully applied this approach to the 1D transverse-field Ising model. It is essential to emphasize that the data presented here is in its raw form, lacking any noise reduction considerations. The summation of noisy data has the potential to degrade the precision of the results. Therefore, we would like to point out two pertinent perspectives: the evaluation of noise reduction techniques on the Score Average and the implementation of a heuristic to navigate in the energy range. These approaches aim to mitigate errors in the NoS estimations and enhance computational efficiency by avoiding prolonged interactions in regions without indications of potential eigenvalues. Given that the quantum system's basis states increase exponentially, we also have in perspective the development of a Monte Carlo method to confront the challenge of studying larger systems.

\section*{Author declarations}

This research received no specific grant from any funding agency in the public, commercial, or not-for-profit sectors.

\subsection*{Conflict of Interest Statement}
The authors have no conflicts to disclose.

\subsection*{Author Contributions}

{\bfseries J.C.S.~Rocha:} Methodology (lead); Writing - Original draft (lead); Writing - review \& editing (lead); Software (lead); Data curation (lead); Formal analysis (lead); Visualization (equal); Conceptualization (supporting).
{\bfseries R.F.I.~Gomes:} Writing – review \& editing (equal); Visualization (supporting); Methodology (supporting); Software (supporting).
{\bfseries W.A.T.~Nogueira:} Writing – review \& editing (equal); Methodology (supporting).
{\bfseries R.A.~Dias:} Conceptualization (lead); Project administration (lead); Visualization (lead); Methodology (equal); Software (equal); Formal analysis (supporting).
\section*{Data Availability Statement}
The data that support the findings of
this study is openly available in
Zenodo at https://doi.org/10.5281/zenodo.10484999, reference number~\citenum{zenodo}.

\appendix
\section{The Score Average of the Rodeo Algorithm} 
\label{appendix_SA}
\begin{figure}[!ht]
\includegraphics[scale=0.4]{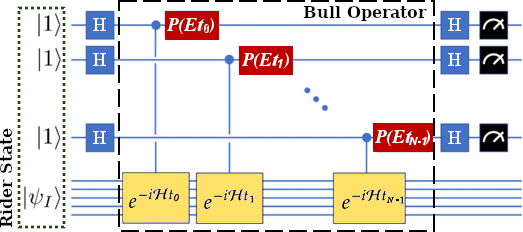}
\caption{Circuit diagram for the rodeo algorithm.}
\label{fig:circuit}
\end{figure}

 To derive a theoretical behavior for the Score Average (SA) of the rodeo algorithm, defined as the average of measurements on the ancillary qubits in the final step of the algorithm, we first introduce the bull operator\footnote{The choice of naming the operator after bull, rather than horse, is attributed to the fact that the former exhibits twisting side-to-side movements, better resembling the rotation of ancillary qubits in the $xy$-plane, in contrast to the up-and-down jumping characteristic of the latter.} and the rider state. The bull operator is defined as the controlled time evolution followed by the phase shift operator, while the rider state is a combined state of the ancillary and the system's qubits, see Fig.~\ref{fig:circuit}.

In a generalized approach, we consider the system described by a rank-$M$ state denoted as $\ket{\psi_I}$. This state is a speculation for the eigenstate of the Hamiltonian, $\mathcal{H}$. Additionally, we also consider that the algorithm is executed by using $N$ ancillary qubits. These ancillary qubits are initially prepared in the state $\ket{1}$. Therefore, initially, the rider state can be expressed as:
\begin{equation}
     \ket{\Psi_0} =  \ket{1}^{\otimes N} \otimes \ket{\psi_I}.
\end{equation}
Subsequently, the ancilla qubits are transformed to the state $\ket{-}$, this is done via the Hadamard tensor. Thus,
\begin{align}  \nonumber 
     \ket{\Psi_1}  &= \left(H^{\otimes N}\otimes \mathds{1}^{\otimes M} \right)\ket{\Psi_0}, \\
     &= \frac{1}{\sqrt{2^N}}   \left[\bigotimes_{k=0}^{N-1}\left(\sum_{y_k=0}^1 e^{i\pi y_k} \ket{y_k}\right) \right] \otimes \ket{\psi_I},
\end{align}
where $\mathds{1}$ is the identity operator.
The bull operator can be constructed as follows: First, the time evolution operator controlled by the $k$-th ancilla can be written as:
\begin{equation}  
\mathcal{CO}_k =  \sum_{x} \Bigg[\mathds{1}^{\otimes (k-1)}\otimes \left( \sum_{y_k=0}^{1} e^{-i E_x y_k t_k} \op{y_k}\right)  \otimes \mathds{1}^{\otimes(N-k)} \Bigg] \otimes  \op{x}.
\label{eq:controlled}
\end{equation}
Here $t_k$ is a random time interval, and $\mathcal{H}\ket{x} = E_x\ket{x}$. Subsequently, the phase shift on the $k$-th ancilla qubit is achieved by the operator:
\begin{equation}
\mathcal{P}_k =  \mathds{1}^{\otimes (k-1)}\otimes  \left( \sum_{y_k=0}^1 e^{i  E t_k y_k} \op{y_k}{y_k} \right) \otimes \mathds{1}^{\otimes (N-k + M)},
\label{eq:phase}
\end{equation}
where $E$ is a guess for the eigenvalue associated with $\ket{\psi_I}$. We can define the operator $\mathcal{B}_k$ that implements the action of the cycle that twists the $k$-th ancilla qubit side to side as:
\begin{align}  \nonumber 
\mathcal{B}_k := &  \mathcal{P}_k\ \mathcal{CO}_k, \\ 
= &\sum_{x} \Bigg[ \mathds{1}^{\otimes (k-1)} \otimes \left( \sum_{y_k=0}^1 e^{i (E - E_x) t_k y_k} \op{y_k}\right) \otimes \mathds{1}^{\otimes (N-k)}\Bigg] \otimes \op{x}. 
\end{align}
By considering that one round is defined when $\mathcal{B}_k$ is performed sequentially on the $N$ ancillary qubits, the bull operator can be expressed as
\begin{align} \nonumber
\mathcal{B}  & := \prod_{k=0}^{N-1}\mathcal{B}_k, \\
&= \sum_{x} \left[ \bigotimes_{k=0}^{N-1} \left( \sum_{y_k=0}^1 e^{i (E - E_{x}) t_k y_k} \op{y_k}\right)\right] \otimes \op{x}.
\end{align}
In this last step, it's important to note that the eigenstates of the Hamiltonian form an orthonormal basis, i.e., $\bra{x'}\ket{x} = \delta[x'-x]$. The rodeo algorithm consists of applying the bull operator to the rider state, i.e.:
\begin{align}\nonumber
\ket{\Psi_i} &= \mathcal{B} \ket{\Psi_1},  \\
&= \sum_{x} \frac{\bra{x}\ket{\psi_I}}{\sqrt{2^N}}  \left[  \bigotimes_{k=0}^{N-1}  \left( \sum_{y_k=0}^1 e^{i [(E -E_{x})t_k + \pi]y_k} \ket{y_k}\right)\right] \otimes \ket{x}.
 \end{align}
 The subsequent step involves changing the basis of the ancillary qubits from the Fourier to the computational basis. This can be achieved by applying the Hadamard tensor again. Thus, the final state can be written as:
\begin{align} 
 \label{final_state}
\ket{\Psi_f} &= \left(H^{\otimes N} \otimes   {\mathrm{1}^{\otimes M}}\right)  \ket{\Psi_i}, \nonumber \\
 &=  \sum_{x}\bra{x}\ket{\psi_I} \Bigg[  \bigotimes_{k=0}^{N-1} \Bigg(  \frac{ 1 -  e^{i (E -E_{x})t_k}}{2} \ket{0} + \frac{ 1 + e^{i (E - E_{x})t_k} }{2} \ket{1} \Bigg)\Bigg]  \otimes  \ket{x}. 
\end{align}
The final step involves measuring the ancillary qubits. The expected value, denoted here by $h(E, \psi_I)$, depends on the measurement process. If the ancillary qubits are simultaneously measured, it leads to:
\begin{align}
\label{he_simul}
 h(E,\psi_I|t) & =   \bra{\Psi_f} \sigma_z^{\otimes N}\otimes \mathds{1}^{\otimes M} \ket{\Psi_f},  \nonumber \\ 
    & =  -\sum_{x} |\bra{x}\ket{\psi_I}|^{2}   \prod_{k=0}^{N-1}\cos{\left[ (E - E_{x})t_k  \right]}.
\end{align}
On the other hand, if the ancillary qubits are sequentially measured, it leads to:
\begin{align}
h(E,\psi_I|t)  & = \frac{1}{N} \sum_{k=0}^{N-1} \bra{\Psi_f} \mathds{1}^{\otimes k}\otimes\sigma_z\otimes \mathds{1}^{\otimes N-k-1+M} \ket{\Psi_f}, \nonumber \\
    & =  -\frac{1}{N}  \sum_{k=0}^{N-1} \sum_{x} |\bra{x}\ket{\psi_I}|^{2}   \cos{\left[ (E - E_{x})t_k  \right]}.
    \label{h_seq}
\end{align}

The SA is the mean value of $h(E,\psi_I|t)$, which depends on the randomly sampled values of $t_k$ drawn by a given probability density function (PDF), denoted as $P(t|\{ X \})$, where $\{X\}$ is a set of parameters that characterize the PDF. We can express the average value as an integral over all possible values of $t$ weighted by the PDF, i.e.:
\begin{eqnarray}
\overline{h}(E,\psi_I)& = \int^{\infty}_{-\infty} h(E,\psi_I|t) P(t|\{ X \}) \mathrm{d}t.
\label{MedGauss}
\end{eqnarray}
To proceed with the calculations, we follow the original recipe by choosing the normal distribution as the PDF:
 \begin{eqnarray}
P(t|\tau,d)& = \frac{1}{d\sqrt{2\pi}} e^{-\frac{(t-\tau)^2}{2d^2}}.
\label{GaussDist}
\end{eqnarray}
For this distribution, $\{X\} = \tau, d$, where $\tau$ is the mean and $d$ is the standard deviation.
To perform this integration, one should realize that:
\begin{equation}
    \frac{1}{d\sqrt{2\pi}}\int^{\infty}_{-\infty} \cos(\alpha t) e^{-\frac{(t-\tau)^2}{2d^2}} \mathrm{d}t  = e^{-\frac{d^2\alpha^2}{2}}\cos(\alpha\tau).
\end{equation} 

Therefore, the SA through simultaneous measurement, i.e., calculated by considering $h(E,\psi_I |t)$ given by Eq.~(\ref{he_simul}), can be written as:
 \begin{equation}
\label{eq:ge}
     \overline{h}(E,\psi_I) = 
     -\sum_{x} |\bra{x}\ket{\psi_I}|^{2}   e^{-\frac{Nd^2\left[ (E - E_{x}) \right]^2}{2}}\cos^N{\left[ (E - E_{x})\tau \right]}.
 \end{equation}
 
Otherwise, for the sequential measurement, where $h(E,\psi_I |t)$ is given by  Eq.~(\ref{h_seq}), leads to:
 \begin{equation}
\label{eq:ge_seq} 
      \overline{h}(E,\psi_I)  = 
     -\sum_{x} |\bra{x}\ket{\psi_I}|^{2}   e^{-\frac{d^2\left[ (E - E_{x}) \right]^2}{2}}\cos{\left[ (E - E_{x})\tau \right]}.
 \end{equation} 
 Based on our simulations, it appears that measurements are conducted in a sequential order. Therefore, we adopt Eq.~(\ref{eq:ge_seq}) as the theoretical relation for the behavior of the SA.

\section*{References}


\bibliographystyle{unsrt}  
\bibliography{DOS_rodeo}

\end{document}